\documentclass[reprint,showpacs,superscriptaddress,twocolumn,english,notitlepage,longbibliography,nofootinbib]{revtex4-1}

\usepackage{amsmath}
\usepackage{amsfonts}
\usepackage{amssymb}
\usepackage[unicode=true,
  bookmarks=true,bookmarksnumbered=true,bookmarksopen=false,
  breaklinks=true,pdfborder={0 0 0},backref=false,colorlinks=true,linkcolor=blue]{hyperref}
\usepackage{cleveref}

\usepackage{graphicx}
\usepackage{bbold}					
\usepackage[dvipsnames]{xcolor}

\usepackage{ulem}
\usepackage{soul,xcolor}

\setstcolor{red}

\usepackage{array,booktabs,ragged2e}	
\usepackage{soul}	
\usepackage{cancel}						
\usepackage{tabularx}					
\usepackage{multirow}
\usepackage{hhline}
\usepackage{adjustbox}
\usepackage{sidecap}

\usepackage[utf8]{inputenc}
\usepackage[T1]{fontenc}

\newcolumntype{P}[1]{>{\raggedleft\arraybackslash}p{#1}}
\newcolumntype{R}[1]{>{\centering\arraybackslash}p{#1}}

\makeatletter
\renewcommand{\p@subsection}{}
\renewcommand{\p@subsubsection}{}
\makeatother

\begin{document}


\title{Magnetoresistance in ZrSi$X$ ($X=$~S, Se, Te) nodal-line semimetals}

\author{ShengNan Zhang}
\email{shengnan.zhang@epfl.ch}
\affiliation{Institute of Physics, Ecole Polytechnique F\'{e}d\'{e}rale de Lausanne (EPFL), CH-1015 Lausanne, Switzerland}
\affiliation{National Centre for Computational Design and Discovery of Novel Materials MARVEL, Ecole Polytechnique F\'{e}d\'{e}rale de Lausanne (EPFL), CH-1015 Lausanne, Switzerland}
\affiliation{Beijing Polytechnic College, Beijing 100042, China}

\author{Oleg V. Yazyev}
\email{oleg.yazyev@epfl.ch}
\affiliation{Institute of Physics, Ecole Polytechnique F\'{e}d\'{e}rale de Lausanne (EPFL), CH-1015 Lausanne, Switzerland}
\affiliation{National Centre for Computational Design and Discovery of Novel Materials MARVEL, Ecole Polytechnique F\'{e}d\'{e}rale de Lausanne (EPFL), CH-1015 Lausanne, Switzerland}
\date{\today}

\begin{abstract} 
We present a comprehensive first-principles study of the magnetoresistance in ZrSi$X$ ($X=$~S, Se, Te) topological nodal-line semimetals. Our study demonstrates that all primary features of the experimentally measured magnetoresistance in these materials are captured by our calculations, including the unusual butterfly-shaped anisotropic magnetoresistance. This anisotropic magnetoresistance can be accurately reproduced using the semiclassical Boltzmann transport theory without introducing any information on the topological nature of bands or the concepts of topological phase transition. Considering the complex structure of the Fermi surface in these topological materials, we develop a theoretical description explaining the features observed in magnetoresistance measurements. Additionally, the atypical Hall resistance can be interpreted by the same semiclassical approach. Our findings establish magnetotransport as a powerful tool for analyzing the geometry of the Fermi surface, complementing angle-resolved photoemission spectroscopy and quantum oscillation measurements. This approach is demonstrated to be particularly useful for determining the role of non-trivial topology in transport properties.
\end{abstract} 

\maketitle
   
\section{Introduction}

Over the past decade, topological materials and their diverse, novel properties have been studied deeply and extensively, among which, magnetoresistance (MR) has garnered significant attention~\cite{Ali2014, Liang2015, Shekhar2015, Gopinadhan2015,Tafti2016, Kumar2017, Zhang2019}. This phenomenon exhibits several intriguing characteristics, such as its presence in a wide range of non-magnetic materials, including both topologically trivial and nontrivial ones, exceptionally large magnitude, lack of saturation even in magnetic fields up to tens of Tesla, and pronounced anisotropy with respect the orientation of the magnetic field. Unraveling the underlying mechanisms of MR could pave the way for groundbreaking advancements in magnetic storage, sensors, switches, and other technological applications.

A prominent example of a material with large non-saturating MR is the type-II Weyl semimetal WTe$_2$~\cite{Ali2014}. The discovery of non-saturating MR in WTe$_2$ has spurred a surge of research on magnetotransport properties of topologically nontrivial semimetals~\cite{Liang2015, Shekhar2015, Gopinadhan2015,Tafti2016, Kumar2017}, which exhibit band crossings between conduction and valence bands in the Brillouin zone (BZ). In three dimensions (3D), band crossings can occur at discrete points, along one-dimensional lines, or loops within the BZ. The former scenario gives rise to Dirac semimetals~\cite{Young2012, Wang2012, wang2013} or Weyl semimetals~\cite{Weyl1929, wan2011, Burkov2011,weng2015, Soluyanov2015}, while the latter results in nodal-line semimetals~\cite{Kim2015}. However, nontrivial band topology may not be the essential factor for large non-saturating MR, as numerous topologically trivial materials are known to exhibit this property~\cite{Mun2012, Takatsu2013, Wang2014, Collaudin2015, Fallah2016, He2016, Yuan2016, Shen2016, Xu2017, Lou2017, Wang2018, Du2018, Lv2018, Chen2020, Zhou2020, Mondal2020, Lu2020}.

Recently, ZrSiS has been investigated through first-principles calculations and angle-resolved photoemission spectroscopy (ARPES) measurements, revealing the existence of a linear band crossing along a loop~\cite{Schoop2016, Neupane2016, Lou2016, Topp2016, Hosen2017, Fu2019}, specifically, the so-called nodal lines protected by non-symmorphic symmetries. Furthermore, its Dirac bands show limear dispersion within a broad range of energies, up to 2~eV above and below the Fermi level, which is significantly larger than those in other Dirac materials, make ZrSiS an accessible and intriguing material to study. Subsequently, ZrSiS was reported to exhibit a large non-saturating MR~\cite{wang2016, lv2016, Ali2016, Singha2017,  Zhang2017, Sankar2017, Pan2018, Leahy2018, Novak2019, Voerman2019,Shirer2019}, followed by its isostructural analogs ZrSiSe and ZrSiTe~\cite{jinhu2016, Shirer2019}. Intriguingly, both ZrSiS and ZrSiSe display butterfly-shaped anisotropic MR, a counterintuitive observation as maximum MR does not coincide with the maximum Lorentz force when the current and magnetic field are perpendicular to each other. 
Several scenarios have been proposed to explain the origin of the large non-saturating and anisotropic MR, such as topological phase transitions~\cite{Ali2016}, angle-dependent effective mass, and quantum lifetimes~\cite{Chiu2019}, but a definitive conclusion is yet to be reached.

While ARPES and quantum oscillation measurements provide a vast amount of information on topological materials, MR measurements remain an indispensable method for investigating the Fermi surfaces of materials. The Boltzmann transport theory facilitates our understanding of magnetotransport properties influenced by the Fermi surface geometry, and vice versa. In this study, we follow the same procedure as in our previous work~\cite{Zhang2019,Zhang2024,Pi2024,Liu2024}, to systematically examine the magnetotransport properties of ZrSi$X$ ($X=$~S, Se, Te) using a combination of first-principles calculations and Boltzmann transport theory within the relaxation time approximation. We first obtain the Fermi surface from density functional theory (DFT) calculations, then solve the Boltzmann equation within the relaxation time approximation to determine the conductivity tensor, and finally provide a comprehensive interpretation of MR in terms of Fermi surface geometry. Our first-principles calculations effectively reproduce the  experimentally observed MR, capturing the main features of both anisotropic MR and Hall resistance in the ZrSi$X$ family of materials. These results establish a comprehensive interpretation of the MR features of ZrSi$X$ materials, surpassing the scope of a simple two-band model fitting or numerous other indirect scenarios. Furthermore, by analyzing the intricate geometry of the Fermi surfaces of these three materials, we achieve full understanding of similarities and differences between the magnetotransport behaviors of ZrSi$X$ ($X=$~S, Se, Te).

Our paper is organized as follows. In Section~\ref{Methodology} we present the details of our computational methodology. Section~\ref{Results} discusses the results for ZrSi$X$ ($X=$~S, Se, Te) compound family. Finally, Section~\ref{Summary} summarizes our work.

\section{Methodology}\label{Methodology}

The conductivity tensor $\pmb{\sigma}$ is calculated in presence of an applied magnetic field by solving the Boltzmann equation within the relaxation time approximation as~\cite{Mermin1976}
\begin{equation}
\pmb{\sigma}^{(n)}(\pmb{B})=\frac{ e^2}{4 \pi^3} \int d\pmb{k} \tau_n \pmb{v}_n(\pmb{k})  \pmb{\bar{v}}_n(\pmb{k}) 
\left(- \frac{\partial f}{\partial \varepsilon} \right)_{\varepsilon = \varepsilon_n(\pmb{k})},
\label{eqn-sigmaij}
\end{equation}
where $e$ is the electron charge, $n$ is the band index, $\tau_n$ is the relaxation time of $n$th band that is assumed to be independent on the wavevector $\pmb{k}$, $f$ is the Fermi-Dirac distribution, $\pmb{v}_n(\pmb{k})$ is the velocity defined by the gradient of a band energy
\begin{equation}
\pmb{v}_n(\pmb{k})=\frac{1}{\hbar} \nabla_{\pmb{k}} \varepsilon_n(\pmb{k}),
\label{eqn-velocity}
\end{equation}
and $\bar{\pmb{v}}_n(\pmb{k})$ is the weighted average of velocity over the past history of the charge carrier
\begin{equation}
\bar{\pmb{v}}_n(\pmb{k}) = \int^0_{-\infty} \frac{dt}{\tau_n} e^{\frac{t}{\tau_n}} \pmb{v}_n(\pmb{k}(t)) .
\label{eqn-aver_velo}
\end{equation}
The orbital motion of charge carriers in applied magnetic field causes the time evolution of $\pmb{k}_n(t)$, written as, 
\begin{equation}
\frac{d \pmb{k}_n(t)}{dt} = - \frac{e}{\hbar} \pmb{v}_n(\pmb{k}(t)) \times \pmb{B}
\label{eqn-evol_k}
\end{equation}
with $ \pmb{k}_n(0)=\pmb{k}$. The trajectory $\pmb{k}(t)$ can be obtained by integrating Eq.~(\ref{eqn-evol_k}). As a consequence, $\bar{\pmb{v}}_n(\pmb{k})$ can be calculated as the weighted average of the velocities  along the trajectory $\pmb{k}(t)$ according to Eq.~(\ref{eqn-aver_velo}). In this semiclassical picture, the Lorentz force does not act on charge carriers since it is perpendicular to $\pmb{v}_n(\pmb{k})$, and therefore energy $\varepsilon_n(\pmb{k})$ remains constant as $\pmb{k}$  evolves in time. This is also evident from Eq.~(\ref{eqn-velocity}), implying that $\pmb{v}$ is normal to the constant energy surface, and consequently $\dot{\pmb{k}}$ 
is tangential to it. Since $\dot{\pmb{k}}$ is also perpendicular to magnetic field $\pmb{B}$, it follows that the $\pmb{k}$ vector traces out an orbit which is a cross-section of the Fermi surface by a plane normal to $\pmb{B}$.

According to the mutual orientation of magnetic field and current, one distinguishes two types of MR---transverse and longitudinal. Since charge carriers are acted upon by the Lorentz force in the field with the direction perpendicular to velocity, we consider only the transverse MR in this work. This is nevertheless sufficient to provide a very rich playground for the comparison with experimental results and for discussing the underlying mechanisms.

Furthermore, we make the relaxation time approximation and neglect interband scattering events and magnetic breakdown. Even though ZrSi$X$ ($X=$~S, Se, Te) are multi-band materials, we only assume one constant relaxation time for all bands. Comparing the results of our calculations with the MR measurements, we find very good agreement. In order to provide a more general view, we plot the results of our calculations as a function of combined variable $B\tau$, which corresponds to a dimensionless quantity $\omega \tau =\frac{e B \tau}{m^*}$. The latter represents a complete revolution of the cyclotron orbit before a carrier is scattered, with $m^*$ being the cyclotron effective mass.
In the case of multi-band systems, such as the semimetals discussed in our work, the total conductivity is the sum of band-wise conductivities, which is then inverted resulting in the resistivity tensor $\hat{\rho} = \hat{\sigma}^{-1}$. In order to analyze the results of calculations for the investigated semimetal systems, we will often plot individual resistivities of electrons and holes.

We have implemented this numerical
approach based on the maximally localized Wannier function
tight-binding model~\cite{marzari1997, souza2001, marzari2012} that was constructed by using
the Wannier90~\cite{mostofi2014} and WannierTools~\cite{wu2018} codes. First-principles calculations reported below have been performed
using the generalized gradient approximation~\cite{perdew1996} as implemented in the VASP code~\cite{kresse1996, kresse1999}. Spin-orbit coupling is not taken into account in the first-principle calculations since it is relatively weak and does not modify the geometry of the Fermi surfaces of the three compounds.

\section{Discussion of results}
\label{Results}
\subsection{ZrSiS}

\begin{figure}
\begin{center}
\includegraphics[width=8.5cm]{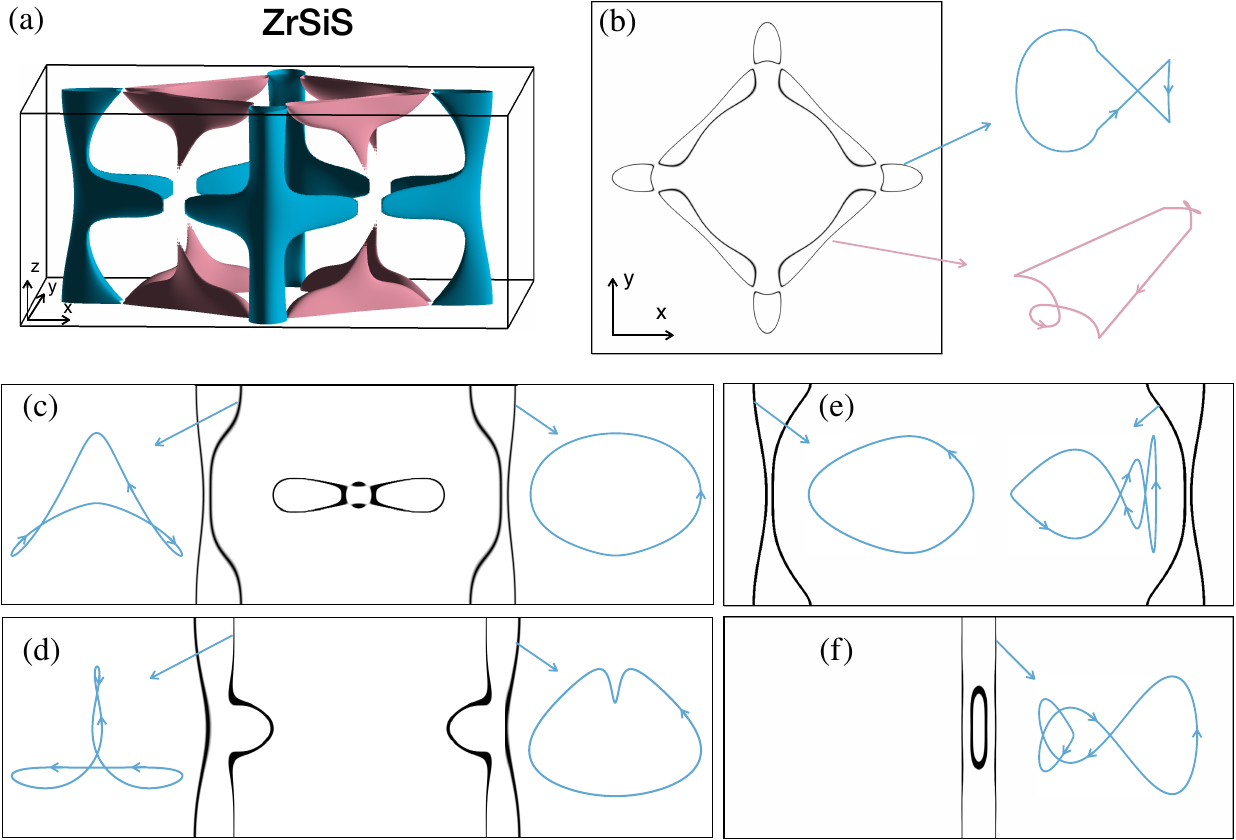}
\caption{(a) Fermi surface of ZrSiS. Blue sheets denote the hole pockets, while the pink ones are electron pockets. The four tube-shaped (blue) Fermi surface sheets extending along the $c$ axis may result in open orbits when magnetic field is applied in the $a$-$b$ plane. (b) A typical cross section of the Fermi surface for B$\parallel$[001] passing through the (0, 0, $\pi$/c) $k$-point. The the blue and pink curves on the right are the corresponding orbits in real space indicated by the same color arrows, respectively. (c),(d) Typical cross sections of the Fermi surface (hole pockets) passing through $k$-points (0.8$\pi$/a, 0, 0) and (0.84$\pi$/a, 0, 0)), respectively, for B$\parallel$[110]. The blue curves are the corresponding orbits in real space. 
(e),(f) Typical cross sections of the Fermi surface (hole pockets) passing through the (0.8$\pi$/a, 0, 0) and (0.84$\pi$/a, 0, 0) $k$-points, respectively, for B$\parallel$[100]. The blue curves show  corresponding orbits in real space.}
\label{fig:fs_zrsis}
\end{center}
\end{figure}

ZrSiS was shown to be a nodal-line semimetal with a series of linearly dispersing Dirac bands at different energies protected by the non-symmorphic symmetry~\cite{Schoop2016}. This gives rise to a diamond-shaped Fermi surface comprising both hole and electron pockets~\cite{Schoop2016, Neupane2016, Lou2016, Topp2016, Hosen2017, Fu2019}. Confirmed by the ARPES measurements~\cite{Fu2019}, the diamond-shaped Fermi surface consists of separate pocket segments: four tube-shaped hole pockets with branches (blue) and four dogbone-shaped electron pockets (pink), as shown in Fig.~\ref{fig:fs_zrsis}(a). These pockets enclose the line of Dirac nodes located near the Fermi energy. Furthermore, quantum oscillation studies have revealed a $\pi$ phase shift in the frequency oscillation pattern, which is associated with the Berry phase acquired by the extremal orbit~\cite{Ali2016}. This is considered an evidence of the existence of the Dirac nodal line enclosed within the Fermi surface. Simultaneously, several groups have observed extremely large positive and non-saturating resistivity, which exhibits strong anisotropy depending on the orientation of the magnetic field~\cite{wang2016, lv2016, Ali2016, Singha2017,  Zhang2017, Sankar2017, Pan2018, Leahy2018, Novak2019, Voerman2019,Shirer2019}.

As demonstrated in the previous studies~\cite{Pippard1989, Chambers1990}, large non-saturating MR occurs if a material exhibits one of two properties: either compensation between hole and electron charge carriers or the presence of open orbits originating from non-closed Fermi surfaces. The geometry of the Fermi surface of ZrSiS is highly complex, as shown in Fig.~\ref{fig:fs_zrsis}(a), with the degree of charge-carrier compensation strongly dependent on the orientation of magnetic field. Additionally, the tube-shaped hole pockets extending along the $c$ direction can generate open orbits under a magnetic field rotating in the $a$-$b$ plane. Based on these observations, we proceed to discuss the intricacies of MR in ZrSiS.

\begin{figure*}[!t]
\begin{center}
\includegraphics[width=17cm]{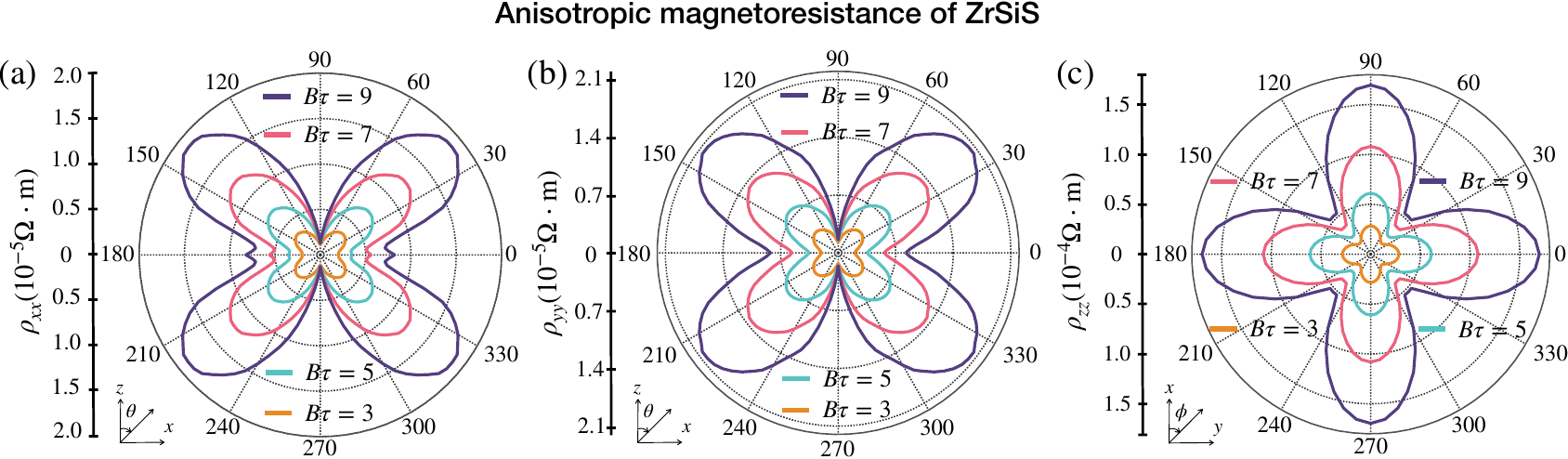}
\caption{
Calculated anisotropic magnetoresistance of ZrSiS for currents along (a) the $a$, (b) $b$ and (c) $c$ axes, and magnetic field rotating in the $b$-$c$, $a$-$c$ and $a$-$b$ planes, respectively. Angles are expressed in degrees ($^{\circ}$). In panels (a) and (b), the maximum resistivity occurs at approximately $\theta = \pi /4  = 45^{\circ}
$ at low magnetic field, and at nearly $\theta = \pi/5  = 36^{\circ}$ at large magnetic field. It has an apparent fourfold symmetry, especially at low magnetic field.}
\label{fig:mr_zrsis}
\end{center}
\end{figure*}

Fig.~\ref{fig:mr_zrsis}(a) displays the polar plot of the anisotropic MR with the current applied along the $a$ axis and magnetic field rotating in the $a$-$c$ plane. The resistivity rapidly increases from a small peak at $\theta = 0$ ($B \parallel c$) to a maximum at around $\theta = \pi /4$, then decreases to another minimum at $\theta = \pi /2$ ($B \parallel a$). The resistivity behavior between $\theta = \pi /2$ and $\theta = \pi$ exhibits the same features, but in reverse order due to the two-fold symmetry of the Fermi surface in the $a$-$c$ plane. The change in MR, with maximum resistivity at $\theta \simeq \pi/4$ (Fig.~\ref{fig:mr_zrsis}(a)), has been reported as the unusual butterfly-shaped MR~\cite{Ali2016, lv2016, jinhu2016, Pan2018, wang2016}. Typically, the MR maximum occurs when current and external magnetic field are perpendicular to each other ($\theta = 0 $), approximately coinciding with the Lorentz force reaching its maximum. However, the maximum resistivity appears at $\theta \simeq \pi/4 $, and its origin remained elusive.

Similarly, the anisotropic MR pattern displayed in Fig.~\ref{fig:mr_zrsis}(b) maintains the butterfly shape when the magnetic field rotates in the $a$-$c$ plane, but with the current changed to run along the $b$ axis, which is consistent with the observed results in Ref.~\cite{Ali2016}. Comparing the two butterfly-shaped anisotropic MR patterns in Figs.~\ref{fig:mr_zrsis}(a) and (b), the smaller (larger) minimum resistivity appearing at $\theta = \pi/2$ ($\theta = 0$) and the maximum resistivity emerging at approximately $\theta = \pi/4$, suggests a possible fourfold symmetry, particularly under low magnetic fields. Given that there are no significant differences in physics between these two scenarios (the magnetic field rotates in the same $a$-$c$ plane), we use the case of current along the $b$ axis to discuss the origin of anisotropic resistivity stemming from the Fermi surface geometry.

First, we examine the butterfly-shaped MR as a function of magnetic-field orientation in terms of symmetry. Due to the two-fold symmetry in the $a$-$c$ plane of the Fermi surface (see Fig.~\ref{fig:fs_zrsis}(a)), the anisotropic MR also shows two-fold symmetry when the magnetic field rotates in this plane. However, the butterfly-shaped anisotropic MR looks as it has an approximately fourfold symmetry, which has been reported by several experimental groups~\cite{Ali2016, wang2016, Novak2019, Chiu2019}. Our calculation and careful analysis suggest that this fourfold symmetry is not rigorous, but rather represents a non-monotonic variation of MR under a magnetic field in the $a$-$c$ plane due to the intricate fine structure of the Fermi surface.

As the magnetic field tilts from the $c$ axis to the $a$ axis in the $a$-$c$ plane, hole carriers gradually change their orbits from closed to open. However, large non-saturating MR induced by open orbits can only be considered if the current is applied along the direction of the open orbit~\cite{Zhang2019}, which is the $c$ axis for ZrSiS. Consequently, the open-orbit mechanism can be ruled out in this case ($I \parallel b$ and $B$ rotated in the $a$-$c$ plane), leaving only the compensation between hole and electron charge carriers to contribute to the resistivity.

Although compensation is the only mechanism to consider, reaching a clear conclusion is challenging due to the complex structure of the Fermi surface. A representative cross-section of the Fermi surface is shown in Fig.~\ref{fig:fs_zrsis}(b) with the magnetic field applied along the $c$ axis. Both electron and hole pockets comprise concave segments, suggesting that the charge carriers originating from the hole (electron) pockets may follow the trajectories of electrons (holes)~\cite{Ong1991}. To identify the type of charge carriers, it is necessary to examine the orbits in real space, which trace the scattering path vector $\pmb{r}(\pmb{k}) = \pmb{v}(\pmb{k}) \tau$ with $\pmb{k}$ evolving on the Fermi surface under a magnetic field. As already illustrated in Ref.~\cite{Ong1991}, the exact shape of the real-space orbits depends entirely on the details of the geometry of the Fermi surface, thus requiring rigorous calculations. In Fig.~\ref{fig:fs_zrsis}(b), we present the real-space orbits corresponding to the Fermi surface cross-sections marked by the arrow with the same color, where the pink one is originates from electron pockets and blue from hole pockets. It is evident that both electron and hole pockets exhibit mixed charge-carrier features in the real-space orbits, which contain clockwise and anticlockwise segments within a single orbit. Consequently, the degree of compensation between the charge carriers depends sensitively on the orientation of the magnetic field.

\begin{figure}[t!]
\begin{center}
\includegraphics[width=8.5cm]{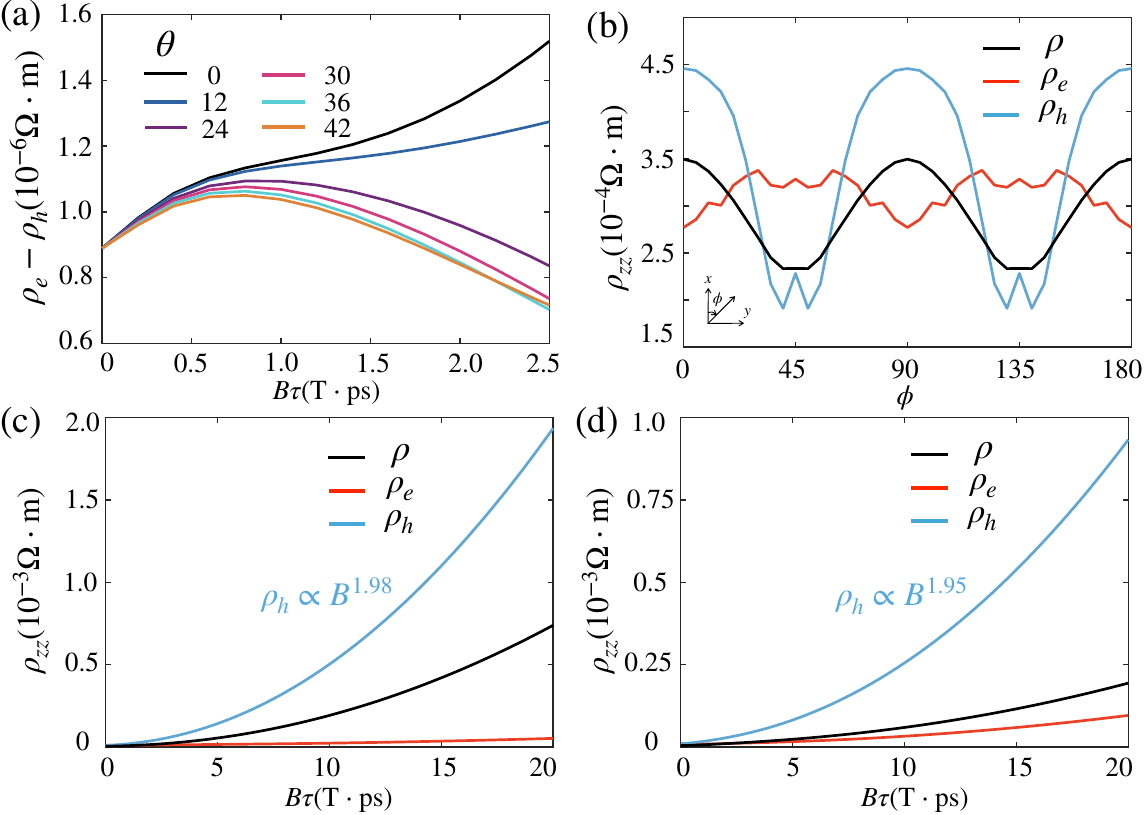}
\caption{(a) Difference between resistivities of electron and hole charge carriers $\rho_{yy}^e - \rho_{yy}^h$ when the magnetic field is rotated from $\theta = 0^{\circ}$ (along the $c$ axis) to $\theta = 42^{\circ}$ (close to the [110] direction).
(b) Resistivity $\rho_{zz}$ as a function of magnetic field
orientation. Field dependences of resistivity $\rho_{zz}$ for magnetic field oriented along (c) the $a$ axis and (d) the [110] direction. The curves in red and blue show individual resistivities $\rho_{zz}$ of electron and hole charge carriers. Individual resistivities of hole charge carriers (blue curve in panels (c) and (d)) show nearly quadratic dependence, indicating that the open-orbit mechanism is dominant.}
\label{fig:mr_zrsis_ana}
\end{center}
\end{figure}

As the magnetic field rotates from the $c$ axis to the $a$ axis, both hole and electron pockets keep producing orbits with concave segments, composited of both clockwise and anticlockwise trajectory segments in real space. Unlike the simple spherical Fermi surface, determining the type of charge carriers for pockets of complex shape is extremely difficult without explicit calculations. Therefore, we calculate the resistivity of the electron and hole charge carriers separately as a function of magnetic field with its orientation changing from the $c$ axis ($\theta = 0$) to the $a$ axis ($\theta = \pi /2$) and plot their difference, $\rho_{xx}^e - \rho_{xx}^h$, in Fig.~\ref{fig:mr_zrsis_ana}(a). The differences gradually become smaller as the magnetic field rotates from the $c$ axis ($\theta = 0$) to the $[110]$ direction ($\theta \simeq \pi /4$), implying that the compensation becomes increasingly efficient, and thus the resistivity reaches its maximum at around $\theta = \pi /4$. This explains the butterfly-shaped anisotropic MR in Fig.~\ref{fig:mr_zrsis}(b).

Furthermore, we would like to point out that the effect of the off-diagonal element of the resistivity tensor is relatively strong in ZrSiS, which means one cannot treat resistivity by simply combining the resistivities of electron and hole charge carriers~\cite{Collaudin2015}, especially under high magnetic fields. Therefore, we only compare the resistivity of electron and hole charge carriers in low magnetic fields in Fig.~\ref{fig:mr_zrsis_ana}(a). According to our calculations, the compensation is most efficient at about $\theta = \pi/4$ (low magnetic field) and $\theta = \pi/5$ (high magnetic field), as shown in Figs.~\ref{fig:mr_zrsis_ana}(a) and \ref{fig:mr_zrsis}(b). We conclude that the butterfly-shaped MR results from the anisotropic electron-hole compensation under magnetic fields due to the unusual shape of the Fermi surface in the $a$-$c$ plane, rather than from the four-fold symmetry of the Fermi surface in the $a$-$b$ plane.

In the following, we will analyze the anisotropic MR shown in Fig.~\ref{fig:mr_zrsis}(c) in relation to the $C_{4z}$ symmetry of the Fermi surface in the $a$-$b$ plane, i.e. for the current applied along the $c$ axis and magnetic field rotated in the $a$-$b$ plane. The resistivity achieves its maximum value when magnetic field is oriented along the $a$ axis ($\phi = 0$) and decreases as magnetic field rotates away from the $a$ axis, reaching its minimum at $\phi = \pi/4$. Furthermore, Fig.~\ref{fig:mr_zrsis_ana}(b) presents the anisotropic MR as a function of magnetic field orientation for electron (red) and hole (blue) charge carriers separately, comparing them with the overall MR (black). The resistivity of the hole charge carriers contributes the most to the MR and shows consistency with the overall MR anisotropy, while the resistivity of the electron charge carriers reveals an entirely opposite dependence, as seen in Fig.~\ref{fig:mr_zrsis_ana}(b). Consequently, the hole charge carriers dominate the magnetotransport under a magnetic field applied in the $a$-$b$ plane.

The presence of a significant number of hole open orbits along the $c$ axis, due to the four tube-shaped hole pockets when the magnetic field rotates in the $a$-$b$ plane, is evident in Figs.~\ref{fig:fs_zrsis}(c) and \ref{fig:fs_zrsis}(d). In addition, these orbits evolve on the Fermi surface with concave segments, resulting in their real-space orbits containing both clockwise and anticlockwise segments, similar to those shown in Fig.~\ref{fig:fs_zrsis}(b). As a consequence, both open orbit and compensation would contribute to the resistivity considering both types of charge carriers. However, our calculations reveal that open orbits originating from the hole Fermi surface play a more significant role in determining the MR, rather than the compensation between electron and hole charge carriers.

In order to provide a comprehensive understanding, we compare the dependence of MR on  magnetic field for electron and hole carriers separately, and for magnetic orientations $B \parallel a$ (Fig.~\ref{fig:mr_zrsis_ana}(c)) and $B \parallel [110]$ (Fig.~\ref{fig:mr_zrsis_ana}(d)). The dependence of MR on magnetic field shows an approximately quadratic scaling in both cases, i.e. $\rho_{h} \propto B^{1.98}$ for $B \parallel a$ and $\rho_{h} \propto B^{1.95}$ for $B \parallel [110]$. 
This is consistent with the open orbits dominating magnetotransport when the magnetic field rotates in the $a$-$b$ plane. In contrast, the resistivity of electron charge carriers reaches saturation quickly as the magnetic field increases. It is important to note that the value of resistivity is highly dependent on the fine details of curvature of the hole pocket, and hence the detailed shape of the orbit in $k$-space. Consequently, it is challenging to provide an intuitive physical picture explaining the origin of the anisotropic MR configuration. Based on our calculations, we conclude that open orbits generated by the magnetic field along the $a$ axis ($\phi = 0$) are more efficient than those along the direction of $\phi = \pi/4$. 
This observation underscores the significance of the open orbit mechanism and the complex Fermi surface geometry in determining the anisotropic MR in materials like ZrSiS.

\begin{figure}
\begin{center}
\includegraphics[width=8.5cm]{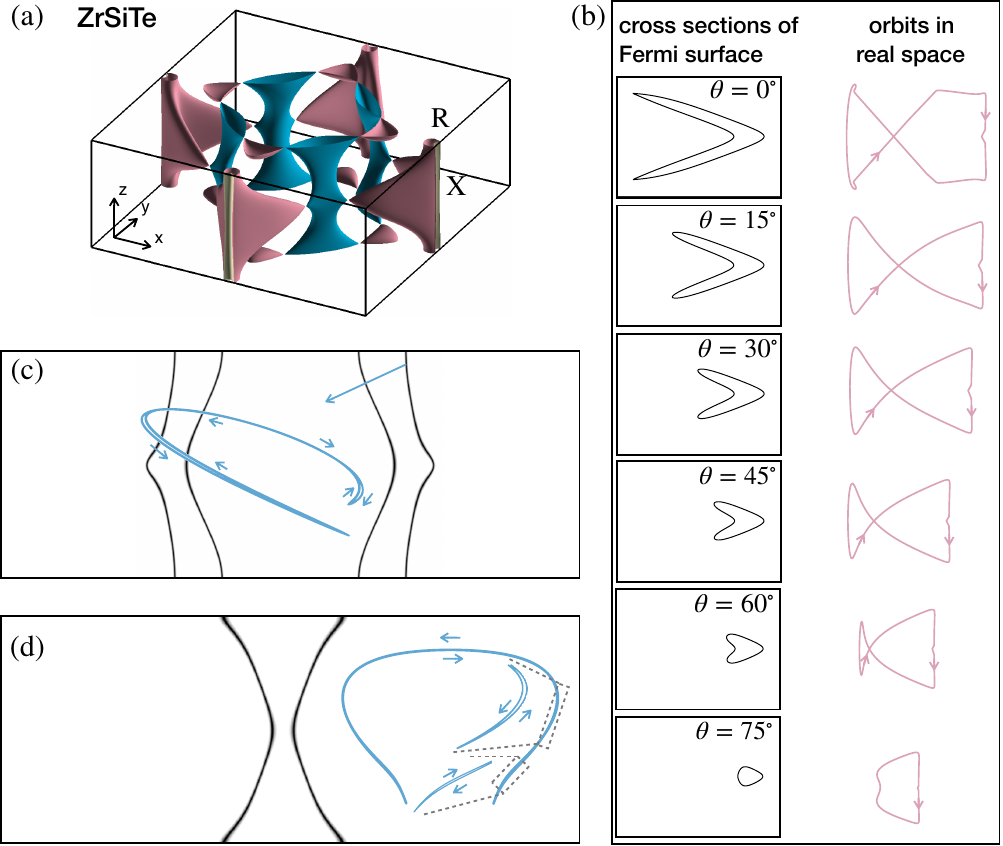}
\caption{(a) Fermi surface of ZrSiTe. Hole and electron pockets are shown in blue and pink, respectively. There are three bands, i.e. one hole band and two electron bands. Although the Fermi surface viewed along the $z$ axis retains the diamond shape, all pockets are significantly different from those in ZrSiS.
(b) Fermi surface cross sections (in black) and the corresponding orbits in real space (in pink). The magnetic field is rotated from $\theta = 0^{\circ}$ (top) to $\theta = 75^{\circ}$ (bottom) by interval of $15^{\circ}$, showing that the content of concave segments is reducing. The arrows mark orientations in which the orbits are followed in the real space, revealing that the content of anticlockwise segments is decreasing as magnetic field is rotating away from the $c$ axis. (c) and (d) Typical cross sections of the Fermi surface of ZrSiTe and the corresponding orbits in real space (blue curves). The cross section in panels (c) pass through points $(0.32\pi/a, 0, 0)$ with $B \parallel [100]$, and in panels (d) pass through points $(0.34\pi/a, 0.34\pi/a, 0)$ with $B \parallel [110]$.
}
\label{fig:fs_zrsite}
\end{center}
\end{figure}

\begin{figure*}[!t]
\begin{center}
\includegraphics[width=12cm]{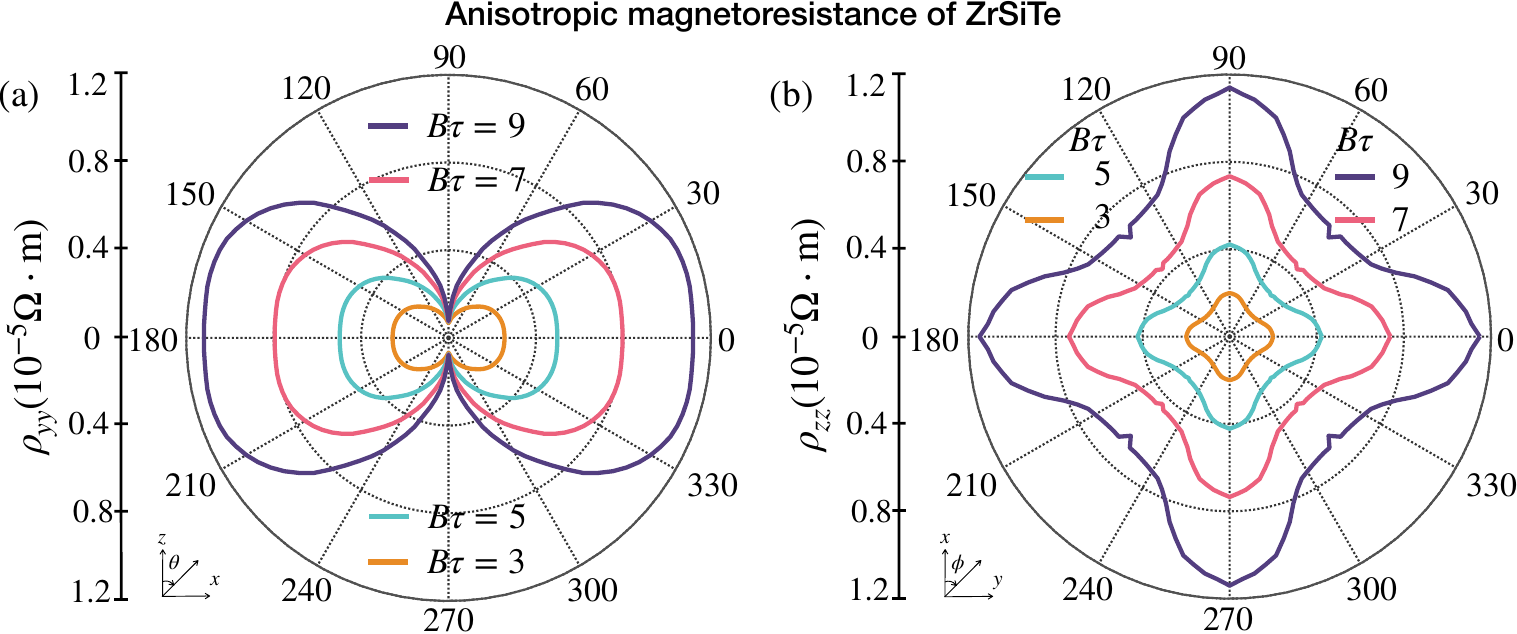}
\caption{Calculated magnetoresistivity anisotropy of ZrSiTe for current oriented along (a) the $b$ and (b) $c$ axes and magnetic field rotated in the $a$-$c$ and $a$-$b$ planes, respectively. MR in panel (a) changes to peanut-shaped in contrast with that of ZrSiS and ZrSiSe while retaining its four-fold symmetry in (b).}
\label{fig:mr_zrsite}
\end{center}
\end{figure*}

\begin{figure}
\begin{center}
\includegraphics[width=8cm]{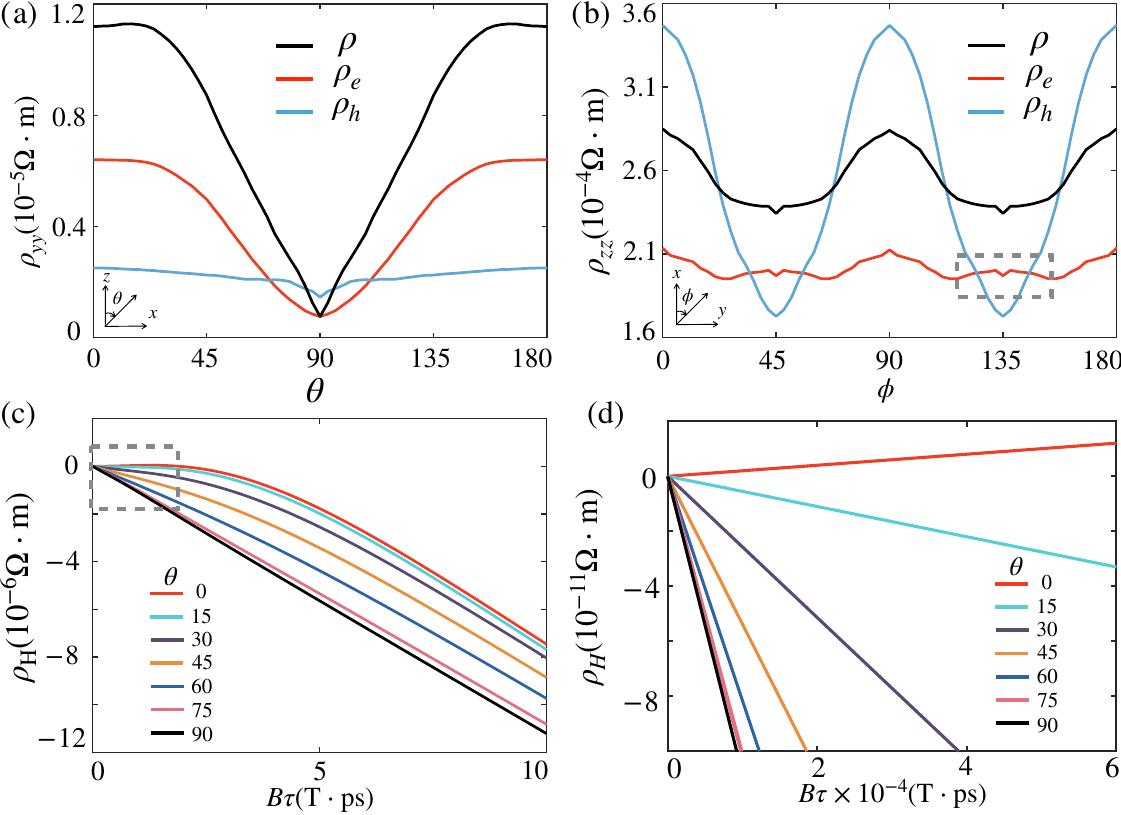}
\caption{(a) and (b) Resistivity $\rho_{zz}$ of ZrSiTe as a function of magnetic field
orientation for current oriented along (a) the $b$ and (b) $c$ axes and magnetic field rotated in the $a$-$c$ and $a$-$b$ planes, respectively.
Individual resistivities $\rho_{zz}$ of electrons and holes are shown in red and blue, respectively. (c) Hall resistivities $\rho_{\rm{H}}$ under different magnetic field orientations in the $a$-$c$ plane (indicated by $\theta$). (d) Details of Hall resistivity $\rho_{\rm{H}}$ near zero magnetic field.}
\label{fig:mr_zrsite_ana}
\end{center}
\end{figure}


\subsection{ZrSiSe}

As ZrSiS and ZrSiSe share similar lattice structures and Fermi surface geometries, their magnetoresistance results are expected to be very similar too. Therefore, we will not delve into a detailed analysis for ZrSiSe, as it would be largely repetitive of the results and discussions already presented for ZrSiS. For further information of ZrSiSe, readers can refer to the Supplemental Material document~\cite{supp}.


\subsection{ZrSiTe}

ZrSiTe shares a similar crystal structure with ZrSiS and ZrSiSe, as reported in Ref.~\cite{Xu2015}, and all three materials are classified as nodal-line semimetals~\cite{jinhu2016}. However, compared to the two already discussed materials, ZrSiTe also shows differences in its band structure and Fermi surface. As shown in Fig.~\ref{fig:fs_zrsite}, both the electron and hole pockets change in their location and shape. In ZrSiTe, the electron pockets involve an additional band, which gives rise to a small tube-like structure enclosing the $R$-$X$ high-symmetry line. Simultaneously, the other electron pocket transforms into a closed section and a tube with four triangle-shaped wings, enclosing an additional small tube-like electron pocket. The hole pockets in ZrSiTe loose their diamond shape in the $a$-$b$ plane, but maintain the open geometry, allowing for open orbits extending along the $c$ axis. These differences in the Fermi surface geometry and band structure of ZrSiTe highlight its unique electronic properties and magnetotransport phenomena when compared to ZrSiS and ZrSiSe.

The calculated anisotropic MR of ZrSiTe with the current applied along the $b$ axis and magnetic field rotated in the $a$-$c$ plane is shown in Fig.~\ref{fig:mr_zrsite}(a). Unlike the butterfly shape observed for ZrSiS and ZrSiSe, the Te-based material exhibits a peanut-shaped anisotropic MR. The resistivity reaches its maximum when the magnetic field is along the $c$ axis and gradually decreases to a minimum when the magnetic field is along the $a$ axis. This qualitative difference in the configuration of the anisotropic MR arises from the discrepancies in the Fermi surface geometries of these materials.  Fig.~\ref{fig:mr_zrsite_ana}(a) presents the individual resistivities as a function of the orientation of magnetic field for electron and hole charge carriers in ZrSiTe. The trend corresponding to the electron charge carriers (red) is similar to that of the combined resistivity, while the hole charge carriers (blue) show a nearly constant resistivity, except at $\theta = \pi /2$. This observation suggests that electron charge carriers dominate charge transport in ZrSiTe. Consequently, it is essential to analyze the behavior of electron charge carriers separately for better understanding of the underlying mechanisms responsible for the observed anisotropic MR in ZrSiTe.

The electron pockets in ZrSiTe are indeed complex, comprising concave segments. As magnetic field rotates from the $c$ axis to the $a$ axis, the type of charge carrier may change. To investigate this possibility, we calculated the Hall resistivity for this electron band separately. The result is shown as a function of magnetic field in Fig.~\ref{fig:mr_zrsite_ana}(c). Under the expectation that charge carriers remain electrons as magnetic field changes direction, the Hall resistivity should be negative. Surprisingly, however, under a magnetic field applied along the $c$ axis ($\theta = 0$), the Hall resistivity initially takes a positive value (Fig.~\ref{fig:mr_zrsite_ana}(d)) before turning negative as the magnetic field increases. This observation suggests that some of the charge carriers originating from the electron pocket behave like holes, while the rest retain their electron nature, resulting in a sign reversal of the Hall resistivity at a certain point. As magnetic field is tilted toward the $a$ axis ($\theta = \pi /2$), the positive value of the Hall resistivity vanishes and remains negative, indicating that electrons begin to dominate the transport. This result highlights the complex nature of electron and hole pockets in ZrSiTe and how the orientation of the magnetic field can influence their contribution to the overall resistivity.

The origin of this intriguing inverted behavior of charge carriers can be understood by analyzing representative cross sections of the Fermi surface (black) and the corresponding real-space orbits (pink) at different magnetic field orientations, as shown in Fig.~\ref{fig:fs_zrsite}(b). As the magnetic field changes from the $c$ axis ($\theta = 0$) to the $a$ axis ($\theta = \pi /2$), not only the entire black curves shrink, but the concave segments within them also reduce in length. Concurrently, five of the six orbits in real space (pink) contain both clockwise and anticlockwise segments, implying that these orbits generate both electron and effective hole charge carriers simultaneously.  Furthermore, the anticlockwise segments in real space shrink together with the concave segments of reciprocal space orbits as the magnetic field rotates from the $c$ axis ($\theta = 0$) to the $a$ axis ($\theta = \pi /2$). This gradual reduction of concave segments of $k$-space orbits diminishes the generation of hole charge carriers. Consequently, the Hall resistivity becomes non-positive when there are more electron charge carriers than effective hole charge carriers originating from the same electron pocket. Eventually, both the concave segments of $k$-space orbits and the anticlockwise segments in real-space orbits disappear (Fig.~\ref{fig:fs_zrsite}(b)). This particular gradual inversion of charge-carrier character results in the resistivity exhibiting two-fold symmetric anisotropy rather than the butterfly-shaped anisotropic MR observed in the case of S- and Se-based counterparts. This unique behavior in ZrSiTe highlights the importance of the Fermi surface geometry in determining the magnetotransport properties and how sensitive these properties can be to the orientation of the applied magnetic field.

When focusing on the resistivity properties under current applied along the $c$ axis and magnetic field rotated in the $a$-$b$ plane, the anisotropic MR in the polar plot, as shown in Fig.~\ref{fig:mr_zrsite}(b), exhibits a four-fold symmetry that reflects the four-fold symmetry of the Fermi surface projected onto the $a$-$b$ plane. The resistivity reaches its maximum at $\phi = 0$ and then drops to its minimum at $\phi = \pi / 4$, completing the variation circle. Similar to the case of ZrSiS (Fig.~\ref{fig:mr_zrsis_ana}(b)), the individual resistivities of hole and electron charge carriers show inconsistent tendencies. Figure~\ref{fig:mr_zrsite_ana}(b) reveals that the resistivity of hole charge carriers (blue) change consistently with the overall resistivity, meaning the resistivity drops in the whole circle from $\phi = 0$ to $\phi = \pi / 4$, while the electron charge carriers exhibit the opposite behavior (see the frame in Fig.~\ref{fig:mr_zrsite_ana}(b)). Therefore, the hole charge carriers dominate the transport when the magnetic field is rotated in the $a$-$b$ plane. Also, similar to the case of ZrSiS, open orbits originating from the hole pockets have concave segments, and thus the orbits in real space are composed of both clockwise and anticlockwise segments, as shown in Figs.~\ref{fig:fs_zrsite}(c) and \ref{fig:fs_zrsite}(d). Following the same analysis as for the case of ZrSiS, we conclude that the maximum of resistivity at $\phi = 0$ is due to the fact open orbits generated by magnetic field along the $a$ axis ($\phi = 0$) are more efficient than the ones along the direction of $\phi = \pi/4$.

\subsection{Hall resistivity}\label{sec:hall}

\begin{figure}
\begin{center}
\includegraphics[width=9cm]{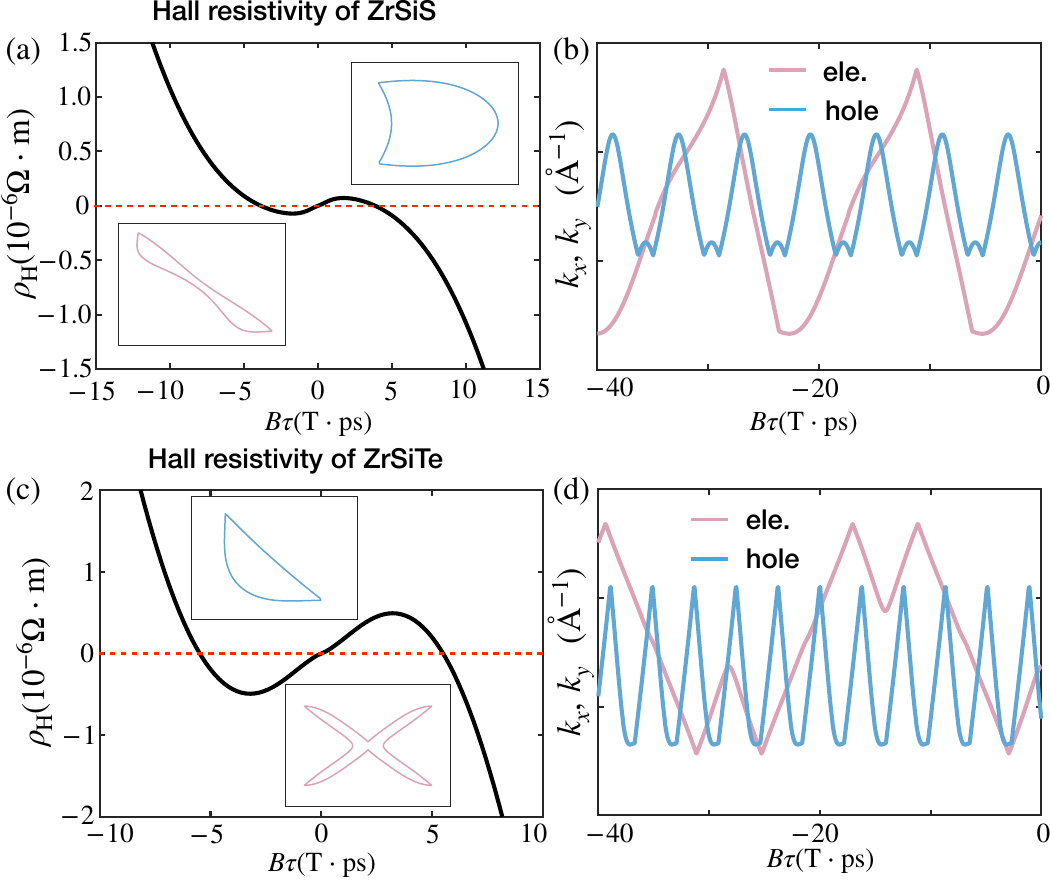}
\caption{(a) Field dependence of the Hall resistivity calculated at magnetic field along the $c$ axis for ZrSiS. The sign reversal takes place at around 4.5~$\rm{T} \cdot \rm{ps}$. The insets in panel (a) show typical hole (blue) and electron (pink) Fermi surface cross sections under magnetic field along the $c$ axis. (b) Examples of cyclotron motion  trajectories in $k$-space driven by magnetic field oriented along the $c$ axis (hole carriers are in blue and electron carriers are in pink). (c) Field dependence of the Hall resistivity of ZrSiTe calculated for magnetic field along $c$ axis. The sign reversal takes place at around 5.5~$\rm{T} \cdot \rm{ps}$. The insets in panel (c) are typical hole (blue) and electron (pink) Fermi surface cross sections under magnetic field along the $c$ axis. (d) Examples of cyclotron motion trajectories in $k$-space driven by magnetic field oriented along the $c$ axis (hole carriers in blue and electron carriers in pink).}
\label{fig:rh_zrsis_te}
\end{center}
\end{figure}

To further study the nature of charge carriers on the Fermi surface, we examine the Hall resistivity of the whole system. We find that the Hall conductance changes sign from positive to negative at low magnetic fields, which is confirmed by experimental Hall measurements~\cite{Singha2017}. This indicates the presence of both electron and hole carriers. The unusual behavior of Hall resistivity can be explained by the cyclotron mass competition of the charge carriers originating from different Fermi surface pockets as the magnetic field increases. To gain a deeper understanding of the nature of charge carriers and Hall resistivity, we analyze the cyclotron motion trajectories in $k$-space driven by magnetic field along the $c$ axis, as shown in Fig.~\ref{fig:rh_zrsis_te}(a). The cyclotron mass can be calculated from the period of the $k$-space trajectories in the time dependence plot in Fig.~\ref{fig:rh_zrsis_te}(b), resulting in $m^*_h = 0.17 m_e$ for the hole carriers and $m^*_e = 0.46 m_e$ for electrons. Considering the fact that the lighter hole carriers complete the cyclotron motion faster than electron carriers (see Fig.~\ref{fig:rh_zrsis_te}(b)), the positive charge carriers first exhibits a positive Hall resistivity slope in small magnetic fields. Later, the negative charge exhibits a negative Hall resistivity slope at larger magnetic fields, as expected. All these features of the Hall conductance exposed in our calculated results are consistent with those observed in experimental measurements~\cite{Singha2017}.

The Hall resistivity behavior of ZrSiTe appears to be quite similar to that of ZrSiS, as shown in Fig.~\ref{fig:rh_zrsis_te}(c), except for a more obvious sign change at around 5.5~T. This phenomenon can be easily understood by examining the cyclotron motion in Fig.~\ref{fig:rh_zrsis_te}(d). In ZrSiTe, the cyclotron mass of the hole carriers decreases to $m^*_h = 0.11 m_e$, while that of the electron carriers increases to $m^*_e = 0.79 m_e$.
As a result, larger (smaller) magnetic fields are needed to allow the heavier (lighter) electron (hole) carriers to complete a period of cyclotron motion in comparison to the lighter (heavier) electron (hole) carriers in ZrSiS. This difference in the cyclotron masses of electron and hole carriers between the two materials leads to the observed differences the bahaviors of Hall resistivity.


\section{Conclusion}
\label{Summary}

In summary, our comprehensive investigation of magnetoresistance and its anisotropy in the family of ZrSi$X$ ($X=$~S, Se, Te) nodal-line semimetals allows us to conclude that these properties can be effectively understood by examining the geometry of the Fermi surface. There is virtually no difference in the geometry of the Fermi surfaces of ZrSiS and ZrSiSe, resulting in highly similar magnetotransport behaviors of these two materials. Both systems exhibit a butterfly-shaped anisotropic magnetoresistance stemming from the non-monotonically varying degree of compensation between the hole and electron charge carriers as the magnetic field rotates, which is due to the intricate details of the Fermi surfaces. In contrast, the geometry of the Fermi surface of ZrSiTe is rather different. As a result, the effective charge-carrier compensation varies monotonically yielding a peanut-shaped anisotropic magnetoresistance rather than the butterfly-shaped one. Additionally, a sign reversal feature in the Hall resistivity reflects the competition between hole and electron charge carriers in these materials. Our study is based on the geometry of the Fermi surface and does not rely on the nontrivial topological properties of the materials, offering insights into the physical mechanisms underlying magnetotransport properties in topologically nontrivial materials, nodal-line semimetals in particular.

\section{Acknowledgments}

We thank Q.S. Wu for invaluable discussions. We acknowledge support by the NCCR Marvel. First-principles calculations have been performed at the Swiss National Supercomputing Centre (CSCS) under Project No. s1146 and the facilities of Scientific IT and Application Support Center of EPFL.


\bibliography{main}
\bibliographystyle{apsrev4-1}   
\end{document}